\documentclass[a4paper]{article}

\usepackage{INTERSPEECH2022}
\usepackage{enumitem}
\usepackage{multirow}

\title{How to Listen? Rethinking Visual Sound Localization}%the Explainable Way}
\name{Ho-Hsiang Wu$^1$, Magdalena Fuentes$^1$, Prem Seetharaman$^2$, Juan Pablo Bello$^1$}
\address{
  $^1$Music and Audio Research Laboratory, New York University, USA \\
  $^2$Descript, Inc.}
\email{hohsiangwu@nyu.edu}

\begin{document}

\maketitle
\begin{abstract}
Localizing visual sounds consists on locating the position of objects that emit sound within an image. It is a growing research area with potential applications in monitoring natural and urban environments, such as wildlife migration and urban traffic. Previous works are usually evaluated with datasets having mostly a single dominant visible object, and proposed models usually require the introduction of localization modules during training or dedicated sampling strategies, but it remains unclear how these design choices play a role in the adaptability of these methods in more challenging scenarios. In this work, we analyze various model choices for visual sound localization and discuss how their different components affect the model's performance, namely the encoders' architecture, the loss function and the localization strategy. Furthermore, we study the interaction between these decisions, the model performance, and the data, by digging into different evaluation datasets spanning different difficulties and characteristics, and discuss the implications of such decisions in the context of real-world applications. Our code and model weights are open-sourced and made available for further applications.
%Furthermore, we propose a simple yet effective approach for localizing sound sources in images and videos, by applying standard contrastive loss for training audio-visual encoders, with no specific localization modules, and then utilizing explainable visualization techniques in the downstream tasks to localize visual objects from audio. We also experiment with different choices of modules including image encoders, training strategies, and visualization techniques to gain insights on the design of such localization systems. 

% We show that this approach performs on par with SOTA (state-of-the-art) models on evaluation datasets consisting of mostly videos with a single dominant source in the image, and outperforms them on a challenging real-world urban traffic dataset. We also show that we can transfer our model to different audio domain (e.g. music) in zero-shot settings and still achieve SOTA performance. Our code and model weights are open sourced and made available for further applications.

\end{abstract}

% \vspace{-1em}
\section{Introduction}
% \vspace{-0.5em}

Audio-visual information is fundamental for the understanding of real-world scenes, as the visual and acoustic characteristics of natural objects provide complementary information that allow us to make sense of them \cite{aytar2016soundnet, arandjelovic2018objects, zhao2018sound, Wang2021_ICASSP}. Modelling the appearance and acoustics of objects not only can help machines understand them better, but also can help them be more efficient in doing so, as shown in recent works which exploit the intrinsic structure of audio-visual data to train models that localize sounding objects without manual labels (self-supervision) \cite{arandjelovic2017look, arandjelovic2018objects, hu2019deep, alwassel2020self, tzinis2021into}. Recently, these audio-visual self-supervised learning techniques became very popular as they have improve their performance in common visual sound localization benchmarks \cite{owens2018audio, senocak2018learning, qian2020multiple, hu2020discriminative}. Most of these proposed approaches require training with carefully designed localization modules (e.g attention, audio-visual fusion) \cite{arandjelovic2018objects, senocak2018learning, harwath2018jointly, tian2018audio, afouras2020self, ramaswamy2020see, chen2021localizing}, and/or dedicated sampling strategies (e.g. hard negative mining) \cite{chen2021localizing, lin2021unsupervised}.

However, there is little understanding on the biases of the audio-visual dataset benchmarks used for visual sound source localization in both training and evaluation of these models, especially when it comes to generalizing to real-world scenarios. Given how challenging it is to annotate and curate such datasets, most of them have either a low number of object categories annotated in few frames and are not suitable for video modeling \cite{senocak2018learning}; have mainly one-dominant object in the whole video \cite{chen2021localizing} which is uncommon in most real-world scenes; or have very particular synthetic visual structure (e.g. objects divided in quadrants) \cite{hu2020discriminative}. To move forward in the localization of sound sources in realistic settings, we think that it is key that we understand better what these models are learning from the data, what design choices are important in these architectures, and what insights our evaluation benchmarks are providing.

Our main contributions are: 1. We analyze various model choices for visual sound localization and discuss how their different components affect the model's performance both quantitatively and qualitatively in different datasets, in particular the encoder' architecture, the loss function and the localization strategy. 2. We provide an extensive analysis of the characteristics of different dataset benchmarks, namely number, area and locations of bounding boxes, audio domains and scene complexity, and discuss their biases in the evaluation of visual sound source localization models. For reproducibility, we open source code and model weights\footnote{github.com/hohsiangwu/rethinking-visual-sound-localization}. %; 2. We propose a simple yet very effective approach for localizing sound sources in images and videos, by applying standard contrastive loss for the training audio-visual encoders, with no specific localization modules, and then using explainable visualization techniques to localize visual objects from audio. We experiment with different choices of image and audio encoders, as well as  explainability visualization techniques, and analyze both quantitatively and qualitatively. 

% 1. We combine both audio-visual self-supervised training and explainability visualization techniques, which allows to use any multi-modal encoder-based models for localization. 2. We experiment with dierse datasets ranging from numbers of bounding boxes, area per boxes, audio domains (e.g. general sound, music, vehicles), and show that we are able to perform on par with SOTA models, while generalizing better to different domains in zero-shot settings. 

% 1. being able to localize sounding objets without any dedicated localization modules, which allows to use any multi-modal encoder-based model for localization as long as it has images. And 2. a more comprehensive and realistic analysis in different datasets with different compositions

% \vspace{-1em}
\section{Design choices for localization}

The different models variations that we study are summarized in Table \ref{tab:model}, along with their main modular design choices. All models exploit a ResNet audio encoder, but they differ on the choice of either image encoder, loss function or localization technique. We explain the different variations in the following.

% We focus our analysis in three main systems. The first one, LVS \cite{chen2021localizing}, consists of 
% To understand the impact of the different modules of the system in the localization we focus

% We hypothesize that audio-visual models trained so that embeddings from different modalities are projected to a joint space is sufficient to localize sounding objects in an image without dedicated localization modules. To that end, as shown in Figure \ref{fig:methods}, we use a contrastive loss for the training of audio-visual encoders, with no specific localization modules, and then using explainablility visualization techniques to localize visual objects from audio. In the following, we explain the different encoders and explainability techniques we study, as well as a comparison to the baselines discussed in Section \ref{subsec:baselines}.

% \vspace{-1em}
\begin{table}[ht!]
\centering
\begin{tabular}{c@{\hskip 0.01in}c@{\hskip 0.01in}c@{\hskip 0.01in}c@{\hskip 0.01in}}
\toprule
Models & Image Enc & Loss & Localization \\
\midrule
\multirow{2}{*}{LVS \cite{chen2021localizing}} & \multirow{2}{*}{ResNet} & Contrastive & Cosine similarity  \\
&  & w/ hard-negatives & w/ sub-patches\\
\midrule
% DSOL \cite{hu2020discriminative} & ResNet & ResNet & custom & N/A \\
% \midrule
RCGrad & ResNet & Contrastive & Grad-CAM \cite{selvaraju2017grad} \\ 
\midrule
% \midrule
% ClipAudioGrad & CLIP & Wav2CLIP & CLIP & Grad \\
CLIPAudioTran & ViT & Contrastive & Transf-MM \cite{chefer2021generic} \\
CLIPTextTran & ViT & Contrastive & Transf-MM \cite{chefer2021generic} \\
\bottomrule
\end{tabular}
\caption{Summary of models and the different design choices. 
}
\label{tab:model}
\vspace{-2em}
\end{table}

% \vspace{-1em}
% \begin{figure}[ht]
% \centering
% \includegraphics[width=\linewidth]{figs/methods.pdf}
% \caption{Our approach: we train the audio and vision encoders using contrastive loss to project both audio and image embeddings into a joint space. We use explainability visualization techniques to obtain heatmaps for object locations.}
% \label{fig:methods}
% \vspace{-1.5em}
% \end{figure}
% We mainly combine two research methods as shown in Figure \ref{fig:methods}: 1. audio-visual contrastive learning as pre-training, 2. utilize explainability visualization techniques on pre-trained image encoder to extract sound source localization information.

\vspace{-0.5em}
\subsection{Audio-visual encoders}
\vspace{-0.5em}

%  we train only with simple contrastive loss on the image and audio level, without the need of specifically designed modules for localization task. We follow closely to the set up of LVS, by taking pre-trained image encoder on imagenet, and randomly initialized audio encoder, trained with VGGSound \cite{chen2020vggsound}, which are youtube videos providing audio-visual correspondence. Our hypothesis is that, with the audio-visual contrastive learning on the clip level, the localization information is already captured/learned in the encoder, which we can extract with explainability visualization methods in Section \ref{subsec:explainbility}.

%We select two different pre-trained audio-visual models as in Figure \ref{fig:methods} on the left, varying in image encoder architectures, and training approaches as listed in Table \ref{tab:model}. And compare these methods with current SOTA models \cite{hu2020discriminative, chen2021localizing}.

% \begin{enumerate}[leftmargin=*]

We use two sets of audio-visual encoders. The first one \textbf{RC}, inspired by LVS \cite{chen2021localizing}, consists of ResNet audio and image encoders, which we initialize randomly for the audio encoder and with ImageNet for the image encoder following \cite{chen2021localizing}. We train this model using contrastive loss on VGGSound. We randomly sample 5-second videos from the training split, use the corresponding audio as input to the audio encoder, and randomly select one image frame from the video.
% Differently from \cite{chen2021localizing}, we do not use image sub-patches in the architecture nor hard-negative mining. %We follow closely \cite{chen2021localizing}, initialize image encoder with pre-trained weights on ImageNet, with audio encoder randomly initialized. We use Adam optimizer with learning rate 0.001, with standard learning rate reduce on plateau, and early stopping. 

The second one \textbf{CLIPAudio} exploits the vision transformer (ViT) \cite{dosovitskiy2020image} from CLIP \cite{radford2021learning}, which was trained with millions of image-text pairs from the internet, and the audio encoder from Wav2CLIP \cite{wu2021wav2clip}, also a ResNet distilled from CLIP via VGGSound with contrastive loss to project audio embeddings into the same space as CLIP's text and image embeddings. We use the pre-trained models from  CLIP\footnote{https://github.com/openai/CLIP} and Wav2CLIP\footnote{https://github.com/descriptinc/lyrebird-wav2clip}.
% We also experiment with text to image localization when text labels are available using only CLIP (ClipText).

%For localizing visual sounds, as shown in Figure \ref{fig:methods} on the right, we forward-pass images through the pre-trained (frozen weights) image encoders, take the activations (or attentions in the case of ViT) from specific layers, compute the gradient of audio embeddings with respect to those chosen layers, and aggregate them to generate heatmaps on the images to indicate localization. We explore two popular visualization methods from the explainability literature:

\vspace{-0.5em}
\subsection{Loss functions}
% \label{subsec:explainbility}
\vspace{-0.5em}

All models use a specific type of contrastive learning, namely InfoNCE \cite{oord2018representation}. The main difference is that LVS \cite{chen2021localizing} compares between audio and image sub-patches w/ hard negative mining for regions within an image \cite{chen2021localizing}, while the other models compare the entire audio and image samples.
% \textbf{Explain briefly both loss functions (1 paragraph max)}

\vspace{-0.5em}
\subsection{Localization techniques}
% \label{subsec:explainbility}
\vspace{-0.5em}
\noindent \textbf{Cosine similarity with sub-patches}: LVS is trained to localize the sound source within the image sub-patches, therefore, the model output consists of localization heatmaps computed using the cosine similarity between the audio and image patches, with interpolation from size 14x14 to actual image size \cite{chen2021localizing}.

% For models trained to contrast on the entire audio and image samples, we can utilize explainability visualization tools to generate the localization heatmaps.

\noindent \textbf{Gradient-based (Grad-CAM \cite{selvaraju2017grad}) (Grad)}: Grad-CAM directly considers the gradients of the loss with respect to the input of each layer, computed and aggregated through back-propagation, in order to generate heatmaps. We use the Grad-CAM implementation\footnote{\text{https://github.com/jacobgil/pytorch-grad-cam}}, which we modify so instead of back-propagating the one-hot encoded class labels, we back-propagate the entire audio embedding as is. We follow the best performing results in \cite{selvaraju2017grad} and select the output of the first normalization layer in the last block of the ResNet encoder as our target layer. For more detail information, please refer to the original paper \cite{selvaraju2017grad}. 

\noindent \textbf{Transformer-based (Transformer-MM \cite{chefer2021generic}) (Tran)}: This approach is designed specifically for transformer models. It combines ideas from both relevancy and attention based explainability methods, by looking at self-attention layers in the encoders. These attention maps of each layer are aggregated recursively through back-propagation in order to generate relevancy maps. We use the implementation\footnote{\text{https://github.com/hila-chefer/Transformer-MM-Explainability}} provided from the authors, and back-propagate both Wav2CLIP audio embeddings and CLIP text embeddings for our study on each modality.

% \subsection{Summary of model variations and benchmarks}

We use the Gradient-based explainability technique with RC because neither encoder is a transformer, and the Transformer-based one with CLIPAudio since in preliminary experiments this technique greatly outperformed Grad-CAM when using ViT \cite{chefer2021generic}. We call these models RCGrad and CLIPAudioTran respectively. To better understand the role of the ViT encoder in CLIPAudio and its match with the Wav2CLIP audio encoder, we include the results of localization using CLIP (CLIPText), i.e. using the image and text encoders for the datasets with available text labels.% (VGGSS and Urbansas).

% \vspace{-1em}
\section{Experimental design}
% \vspace{-0.5em}

%In this section, we discuss the experimental design including downstream localization datasets, model variations, implementations for explainability techniques, and evaluation metrics.

\vspace{-0.5em}
\subsection{Evaluation datasets}
\vspace{-0.5em}
We focus our analysis in four different datasets for sound source localization, spanning different sound sources and localization difficulties, as depicted in Table \ref{tab:dataset}.% and Figure \ref{fig:dist}. 

% \vspace{-1em}
\begin{table}[htb]
\centering
\begin{tabular}{c@{\hskip 0.07in}c@{\hskip 0.07in}c@{\hskip 0.07in}c@{\hskip 0.07in}c@{\hskip 0.07in}c@{\hskip 0.07in}}
\toprule
Dataset & \# Frames & \# BBoxes & Audio Len & Domain & Labels \\
\midrule
FS \cite{senocak2018learning} & 250 & 1 & 20s & General & N \\
VS \cite{chen2021localizing} & 4436 & 1-5 & 10s & General & Y \\
MS \cite{hu2020discriminative} & 455 & 2 & 1s & Music & N \\
US \cite{fuentes2022urbansas} & 10802 & 1-25 & 1s & Vehicle & Y \\ % 15853
\bottomrule
\end{tabular}
\caption{Evaluation datasets: Flickr-SoundNet (FS), VGGSS (VS), Music Synthesis (MS), and Urbansas (US).% with various number of samples, number of bounding boxes (BBoxes) per image, input audio length, and audio domains.
}
\label{tab:dataset}
\vspace{-2em}
\end{table}

% \begin{enumerate}[leftmargin=*]
\noindent \textbf{Flickr-SoundNet (FS)} \cite{senocak2018learning} testset is a popular benchmark for visual sound source localization. It contains 250 image audio pairs with 20s audio with 3 annotation bounding boxes from different annotators. We follow the same pre-processing steps as \cite{chen2021localizing} and take overlapped region agreed on from at least two annotators as ground-truth.

\noindent \textbf{VGGSS (VS)} \cite{chen2021localizing} is a subset from VGGSound testset, labeled with more than one bounding boxes per each image of 220 classes. We are able to acquire 4436 10-second videos from YouTube, and we use the center image frame as originally proposed in \cite{chen2021localizing}.
%, using the entire 10s audio as inputs. 

\noindent \textbf{Music Synthesis (MS)} \cite{hu2020discriminative} is a synthetic audio-visual music dataset. Each audio-visual pair is constructed by concatenating four music instrument frames and randomly selecting two out of four to be the ones producing sound, with corresponding 1s audio, resulting in 455 pairs.

\noindent \textbf{Urbansas (US)} \cite{fuentes2022urbansas} is a new dataset released with real-world urban traffic scenes, featuring vehicle labels including car, motorcycle, bus and truck. It is a video dataset of 10-second clips and stereo audio, with video annotations at 2fps with variety of bounding boxes number ranging from 1 to 25, along with sound events annotations in the audio. We sampled audio-image pairs of 1s audio with the image centered, and we only take those samples with labeled bounding boxes overlapping with audio annotations, resulting in 10802 image frames.
% \end{enumerate}

\begin{figure*}[ht]
\centering
\minipage{0.5\textwidth}
  \includegraphics[width=\linewidth]{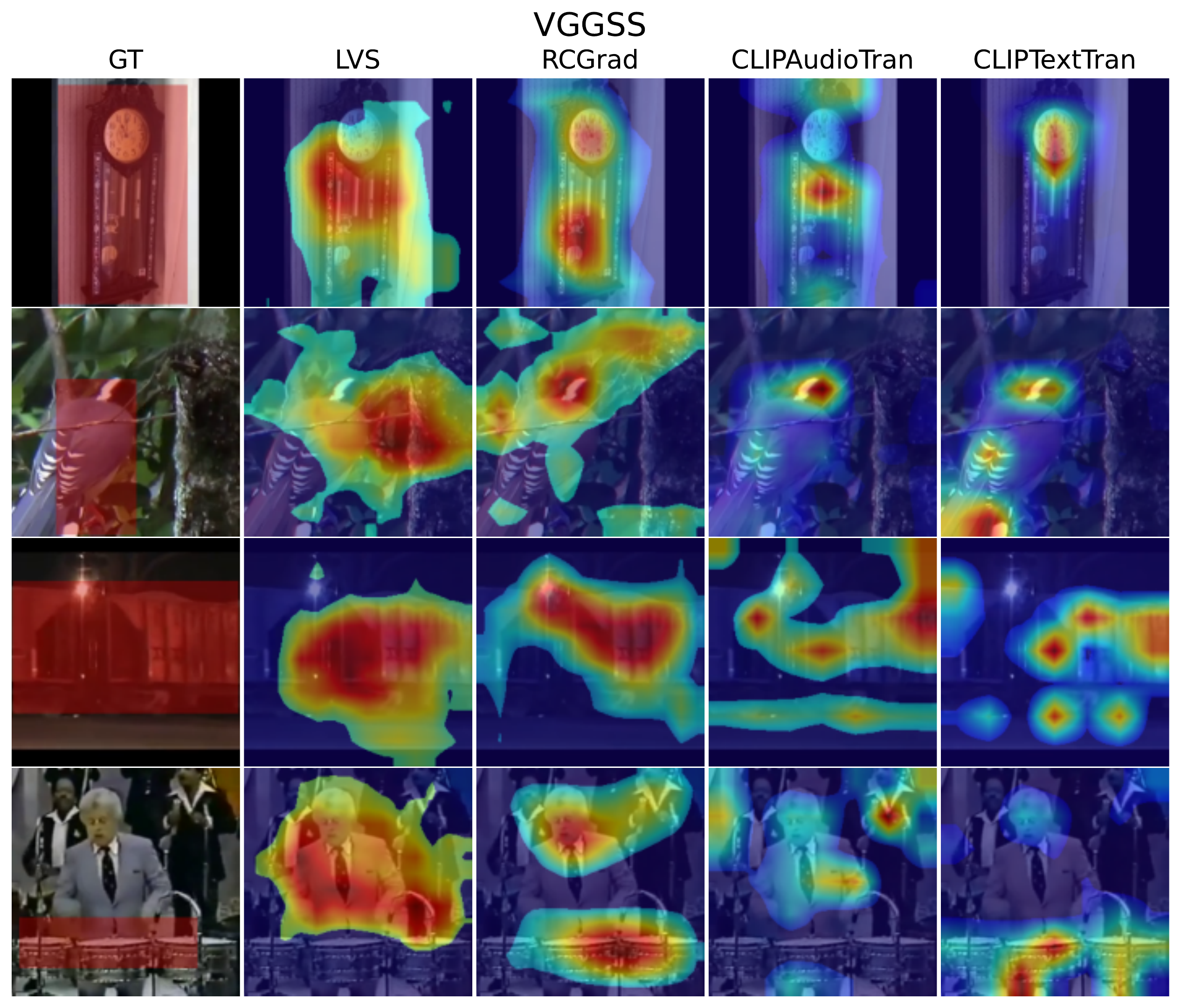}
\endminipage
% \minipage{0.33\textwidth}
%   \includegraphics[width=\linewidth]{figs/quali_ms.pdf}
% \endminipage
\minipage{0.5\textwidth}
  \includegraphics[width=\linewidth]{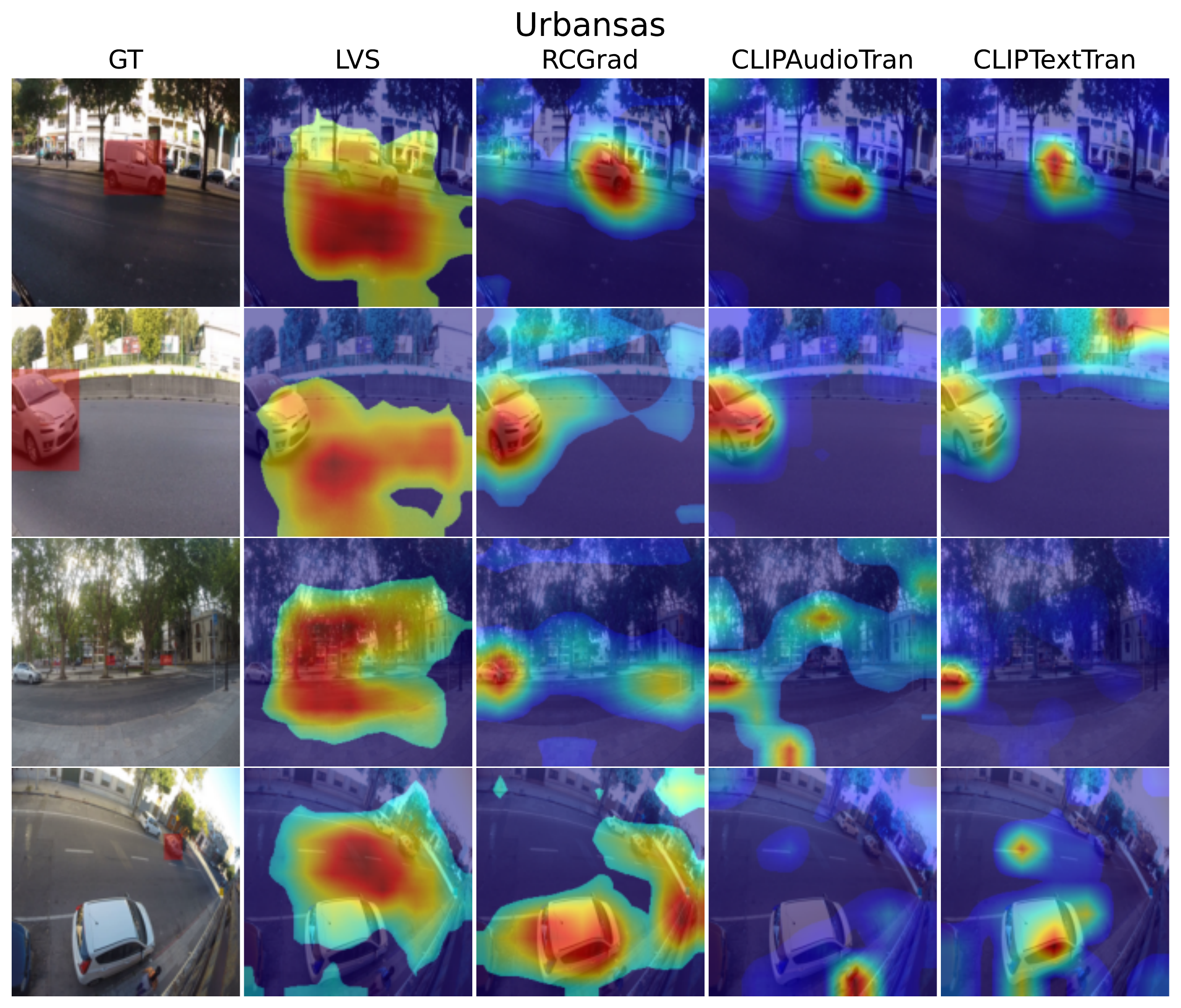}
\endminipage
\caption{Examples of different models (LVS, RCGrad, CLIPAudioTran, CLIPTextTran) performance on VGGSS and Urbansas. GT is ground-truth. Top two rows are examples where models perform better on average, bottom two rows are worse on average.}
\label{fig:quali}
\vspace{-2em}
\end{figure*}

\vspace{-0.5em}
\subsection{Baselines}
\vspace{-0.5em}

\begin{figure}[ht]
\centering
\includegraphics[width=\linewidth]{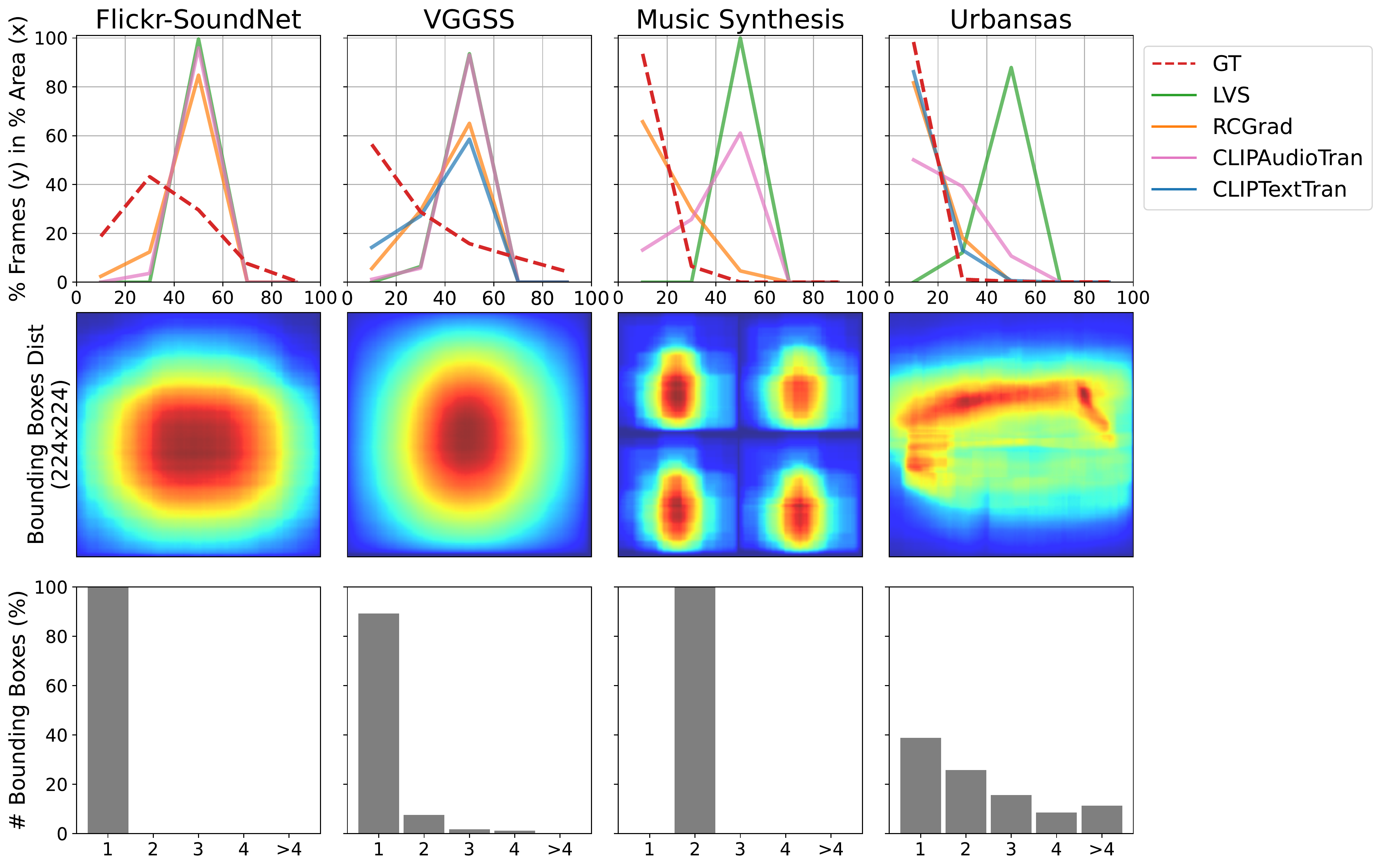}
\caption{Top row is the distribution (in percentage number of frames) of the bounding boxes' area (red dotted line) and predicted heatmap area (solid color for different models) percentage to the entire image. Middle row is the location distribution of each bounding box in the resized 224x224 images. Bottom row is the number of bounding boxes per frame in different datasets.}
\label{fig:dist}
\vspace{-2em}
\end{figure}

Figure \ref{fig:dist} shows the characteristics of each dataset, these statistics illustrate the type of scenes that compose the different datasets. Flickr-SoundNet and VGGSS have most of their bounding boxes located in the middle of the image and occupying a considerable portion of the image, with more than half of the frames covering over 30\% of it. These two datasets are totally (FS) or mostly composed (89.3\%, VS) of clips with a single bounding box (i.e. one sounding object at a time). Inspired by this, we think that a good naive baseline for these datasets would be one bounding box centered in the middle of the image covering at least 50\% of it.

In contrast, all examples in Music Synthesis have at least two sources, and bounding boxes are considerably smaller than the previous two datasets, with most of them being less than 30\% of the image, and their centers are very precisely located in one of four quadrants in the image. We hypothesize that a good baseline in this case would be to locate four boxes in four of those quadrants spanning 1/8 of the image, since it would consistently hit half of the ground-truth positions. Finally, Urbansas bounding boxes are also almost always smaller than 30\% of the image, but their location is very difficult to predict naively, since the dataset comprises images of moving vehicles that constantly move from side to side of the image and thus the center of the bounding boxes not concentrated in any particular section. Urbansas also has different amounts of sounding vehicles per scenes, with the majority of the dataset having one or two. Given that there is no clear naive baseline for this dataset, we use a bounding box centered in the middle covering 50\% of the image, which will cover most cases with multiple vehicles and their location in the scene.

The intuition behind these baselines is to help us understand the difficulties of each dataset as well as how much improvement do the models get from the simplest things we could do with the data. Besides these baselines, we compare with the state-of-the-art (SOTA) models for Flickr-SoundNet and VGGSS \cite{chen2021localizing} (LVS), trained with specific localization and hard negative mining, and for Music Synthesis \cite{hu2020discriminative} (DSOL), trained on music domain data specifically, and require object-level representations. We use the LVS model\footnote{github.com/hche11/Localizing-Visual-Sounds-the-Hard-Way} provided by the authors, pre-trained with VGGSound \cite{chen2020vggsound}, an audio-visual YouTube video dataset containing $\sim$200k 10-second clips (16kHz sample rate) labeled with 309 multi-classes, and follow \cite{chen2021localizing} for the implementation to output localization heatmaps. % and report metrics on the subset of VGGSS we are able to acquire.

\vspace{-0.5em}
\subsection{Evaluation metrics}
\vspace{-0.5em}

Following \cite{senocak2018learning, chen2021localizing}, we apply the same post-processing of output heatmap predictions by min-max normalization, outputting heatmaps with values between 0 and 1. For metrics, we compute the consensus intersection over union (cIoU), namely cIoU@0.5 for Flickr-SoundNet and VGGSS following previous works, meaning that we binarize the heatmap predictions with a threshold equal to 0.5. For Music Synthesis, we use cIoU@0.3 in order to compare with \cite{senocak2018learning}, and we also apply cIoU@0.3 for Urbansas. We report the area under curve (AUC) as well, to do so, we compute the IoU with different thresholds.

\section{Results and discussion}
% \vspace{-0.5em}

\begin{table*}[htb]
\centering
\begin{tabular}{ccccccccccccc}\toprule
& \multicolumn{2}{c}{Flickr-SoundNet (FS)} & \multicolumn{2}{c}{VGGSS (VS)} & \multicolumn{2}{c}{Music Synthesis (MS)} & \multicolumn{2}{c}{Urbansas (US)} \\
% \cmidrule(lr){2-3} \cmidrule(lr){4-7} \cmidrule(lr){8-11}
% & \multicolumn{2}{c}{Audio} & \multicolumn{2}{c}{Audio} & \multicolumn{2}{c}{Text} & \multicolumn{2}{c}{Audio} & \multicolumn{2}{c}{Audio} & \multicolumn{2}{c}{Text} \\
\cmidrule(lr){2-3} \cmidrule(lr){4-5} \cmidrule(lr){6-7} \cmidrule(lr){8-9}
& cIoU@0.5 & AUC & cIoU@0.5 & AUC & cIoU@0.3 & AUC & cIoU@0.3 & AUC \\
\midrule
% Baseline & 0.618 & 0.534 & \textbf{0.323} & \textbf{0.372} & 0.053 (0.022) & 0.195 (0.166) & 0.009 (0.011) & 0.070 (0.074) \\
% R+L \cite{chen2021localizing} & 0.667 & 0.549 & 0.263 & 0.345 & 0.044 & 0.144 & 0.006 (0.008) & 0.065 (0.071) \\
% R+CT+G & \textbf{0.679} & \textbf{0.551} & 0.172 & 0.334 & \textbf{0.369} & \textbf{0.255} & \textbf{0.046} (0.050) & \textbf{0.143} (0.149) \\
Baseline & 0.618 & 0.534 & \textbf{0.323} & \textbf{0.372} & 0.053 & 0.195 & 0.038 & 0.075 \\
\midrule
LVS \cite{chen2021localizing} & 0.667 & 0.549 & 0.263 & 0.345 & 0.044 & 0.144 & 0.036 & 0.071 \\
% LVS \cite{chen2021localizing}& \textbf{0.735}*  & \textbf{0.590}*  & \textbf{0.344}*  & \textbf{0.382}*  & 0.044 & 0.144 & 0.036 & 0.071 \\
DSOL \cite{hu2020discriminative} & - & - & - & - & 0.323*  & 0.235*  & - & -\\
\midrule
RCGrad & \textbf{0.679} & \textbf{0.551} & 0.226 & 0.334 & \textbf{0.369} & \textbf{0.255} & 0.175 & 0.145 \\
\midrule
%ClipAudioGrad & 0.084 & 0.290 & 0.067 & 0.210 & 0.026 & 0.144 & 0.028 & 0.064 \\
% V+CL+G & 0.084 & 0.290 & 0.049 & 0.210 & 0.026 & 0.144 & 0.002 (0.004) & 0.061 (0.061)  \\
% VCG (Text) & & & 0.058 & 0.211 & & & 0.037 (0.046) & 0.128 (0.141) \\
CLIPAudioTran & 0.482 & 0.491 & 0.225 & 0.316 & 0.064 & 0.150 & 0.073 & 0.091 \\
% V+CL+T & 0.482 & 0.491 & 0.226 & 0.316 & 0.064 & 0.150 & 0.011 (0.014) & 0.089 (0.093) \\
%\midrule
CLIPTextTran & - & - & 0.241 & 0.322 & - & - & \textbf{0.219} & \textbf{0.164} \\
% \midrule
% SOTA & \textbf{0.735} \cite{chen2021localizing} & \textbf{0.590} \cite{chen2021localizing} & \textbf{0.344} \cite{chen2021localizing} & \textbf{0.382} \cite{chen2021localizing} & 0.323 \cite{hu2020discriminative} & 0.235 \cite{hu2020discriminative} \\
% SOTA & \textbf{0.747} \cite{lin2021unsupervised} & \textbf{0.596} \cite{lin2021unsupervised} & & & 0.323 \cite{hu2020discriminative} & 0.235 \cite{hu2020discriminative} \\
\bottomrule
\end{tabular}
\caption{Results for different models across datasets. Results indicated with * are reported by the respective authors.} %SOTA for FS and VS are quoted from the original LVS paper for reference, while the row of LVS is computed with exact same implementation as LVS on our acquired subset.}
\label{tab:results}
\vspace{-2em}
\end{table*}

%In addition, Figure \ref{fig:model_pred} shows the area distribution for the bounding boxes from each dataset in red dotted line, and in solid colors the predicted heatmaps from our models and LVS \cite{chen2021localizing} whose code and model weights are publicly available. 

Results are presented in Table \ref{tab:results}. Note that the LVS results (FS and VS) in Table \ref{tab:results} are different from the original paper \cite{chen2021localizing}, which could be due to slightly different data. 

% \textbf{Describe results on Table 2 and highlight that among audio-visual models, RCCGrad does better in all datasets, LVS does good FS and VS, and ClipAudioTran is underperforming compared to RCGrad}

\vspace{-0.5em}
\subsection{Qualitative analysis: what are the models localizing?}
\vspace{-0.5em}

% \begin{itemize}
%     \item Describe results on Table 2 and highlight that among audio-visual models, RCCGrad does better in all datasets, LVS does good FS and VS, and ClipAudioTran is underperforming compared to RCGrad
%     \item Discuss Figure 1 pointing out what each model tends to learn (LVS, big patches in image, ClipAudioTran small regions, RCCGrad good in general, etc)
% \end{itemize}

Figure \ref{fig:quali} shows some qualitative examples of the model predictions. We observe that LVS tends to predict big regions in the center of the image regardless of the scene, RCGrad tends to adapt better across scenes, and CLIPAudioTran tends to predict smaller heatmap regions than the others (top two rows), and predict larger regions when the scenes are challenging (bottom two rows). For instance, in the first two rows under VS, CLIPAudioTran seems to predict the actual sounding part (chimes in the clock, head of a bird) instead of the entire object, causing IoU to drop. This is an interesting observation that distillation from CLIP might actually attend to more granular sound sources.

% As reflected in the results and in Figure \ref{fig:model_pred}, RCGrad performs better in Music Synthesis, consistently locating the instruments in the four quadrants, with some mistakes in difficult camera angles.  

% instead of localizing to the entire object (bird), ClipAudioTran focuses on the head, 

The third row under US in Figure \ref{fig:quali}, shows an example where the models under-perform, this happens particularly when the objects are very small in the scene, and when there are other background noises interfering (e.g. wind or background crowds). In these challenging situations, both RCGrad and CLIPAudioTran tend to also focus on other regions of the image. We hypothesize that this could improve by pre-processing the audio to enhance differences between foreground versus background sounds \cite{lostanlen2018per}. An interesting example is depicted in the last row of Figure \ref{fig:quali} under US. In this case, there are two vehicles in the scene, one is parked and the other one is moving, and the latter is producing sound and thus is the one annotated. Both RCGrad and CLIPAudioTran focus on the larger vehicle parked by the street, which would be a mistake in a monitoring system. This prediction makes sense since the models do not have any temporal context, and we will explore in future work relying on the sequence of images instead of independent frames, and combining with multi-channel audio in order to exploit phase differences indicative of the objects position and motion.

 %are both leaning to the smaller side, mostly smaller than 10\% of the entire image size, and also the bounding box is less concentrated in the center region, with more varieties which match better to the real-world applications. The bottom row shows the number of bounding boxes for each dataset, Urbansas comes with the most diverse of bounding box area and number of boxes.

\vspace{-0.5em}
\subsection{Looking deeper: baselines and datasets distributions}
\label{subsec:baselines}
\vspace{-0.5em}

\begin{figure}[ht]
\centering
\includegraphics[width=\linewidth]{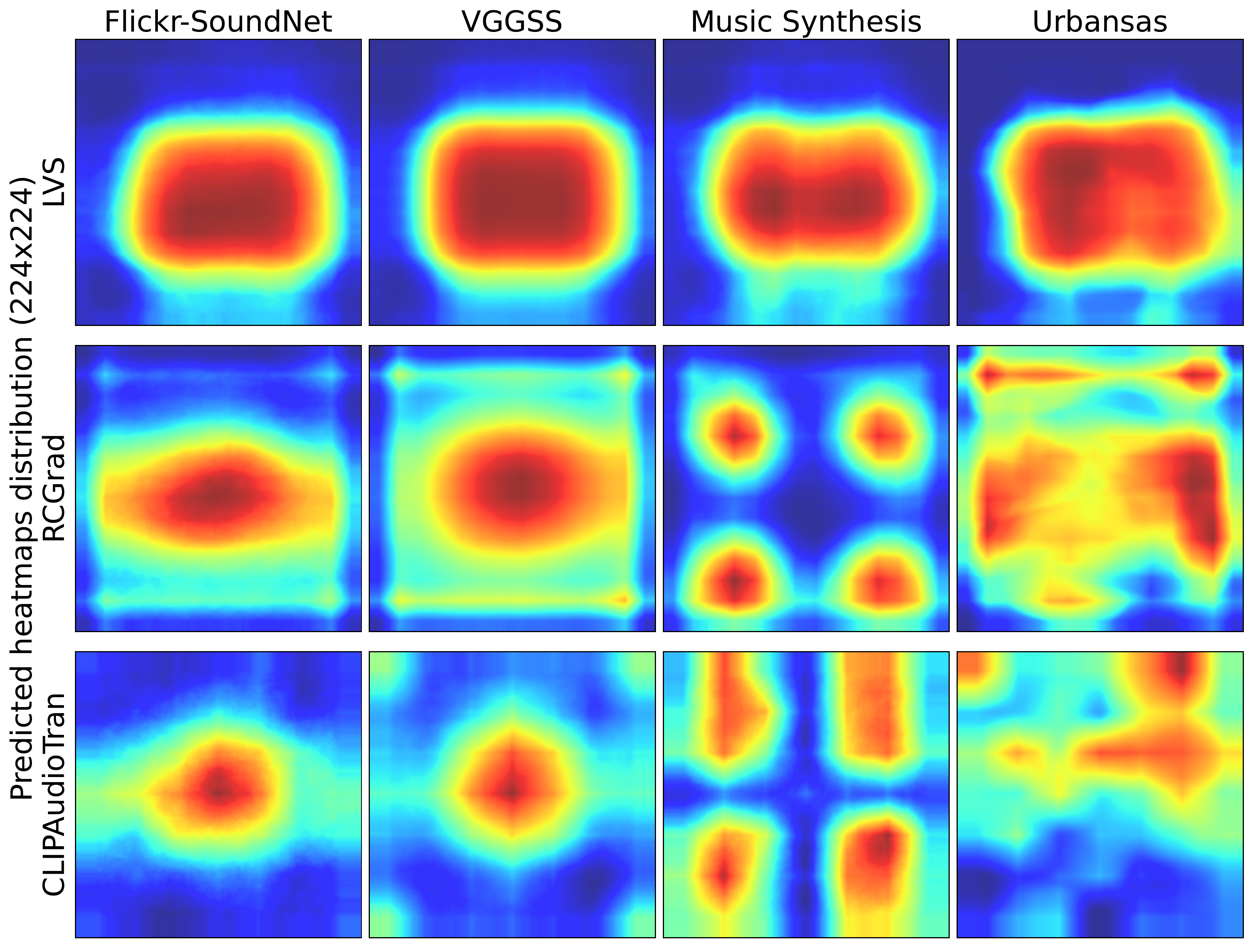}
\caption{Distribution of predicted heatmaps for each model on each dataset. 224x224 image each.}
\label{fig:preds_dist}
\vspace{-1.5em}
\end{figure}

% Figure \ref{fig:dist} shows, for each dataset, the distribution of bounding boxes' sizes (top row), a scatter plot of their locations in the resized 224x224 image (middle row), and the number of bounding boxes per frame (bottom row). These statistics illustrate the type of scenes that compose the different datasets. 

% Flickr-SoundNet and VGGSS have most of their bounding boxes located in the middle of the image and occupying a considerable portion of the image, with more than half of the frames covering over 30\% of it. These two datasets are totally (FS) or mostly composed (89.3\%, VS) of clips with a single bounding box (i.e. one sounding object at a time). Inspired by this, we think that a good naive baseline for these datasets would be one bounding box centered in the middle of the image covering at least 50\% of it. 

Looking at both the actual distribution of predicted heatmaps in Figure \ref{fig:preds_dist}, and middle row of Figure \ref{fig:dist}, we can see that LVS always predict around 50\% of area regardless of the datasets, which fits well with the bounding box area distribution from FS and VS, indicating also why we see good performance from our naive baseline models by predicting a box at the center with 50\% of the image's area. In the case of MS and US, the objects appear in smaller areas in the image, so the results of LVS are worse, even than the naive baselines as in Table \ref{tab:results}.

% In contrast, all examples in Music Synthesis have at least two sources, and bounding boxes are considerably smaller than the previous two datasets, with most of them being less than 30\% of the image, and their centers are very precisely located in one of four quadrants in the image. We hypothesize that a good baseline in this case would be to locate four boxes in four of those quadrants spanning 1/8 of the image, since it would consistently hit half of the ground-truth positions. Finally, Urbansas bounding boxes are also almost always smaller than 30\% of the image, but their location is very difficult to predict naively, since the dataset comprises images of moving vehicles that constantly move from side to side of the image and thus the center of the bounding boxes not concentrated in any particular section. Urbansas also has different amounts of sounding vehicles per scenes, with the majority of the dataset having one or two. Given that there is no clear naive baseline for this dataset, we use a bounding box centered in the middle covering 50\% of the image, which will cover most cases with multiple vehicles their location in the scene. 

Among all models, RCGrad is the one that adapts better to the different datasets, as can be seen by how the area of the predictions matches more closely the ground-truth in Figure \ref{fig:dist} (middle row) and \ref{fig:preds_dist} for most cases, especially MS and US. That also is reflected in its performance in those datasets as in Table \ref{tab:results}. CLIPAudioTran tends to produce predictions that cover big areas when it is uncertain, and though it seems to work well in FS and VS, there is a big gap in performance between this model and the rest as in Table \ref{tab:results}, which suggests that the predictions are focusing on the actual sounding area when the model does well and are likely miss-placed when the model is uncertain, both causing IoU to drop.

% The intuition behind these baselines is to help us understand the difficulties of each dataset as well as how much improvement do the models get from the simplest things we could do with the data. Besides these baselines, we compare with the state-of-the-art (SOTA) models for Flickr-SoundNet and VGGSS \cite{chen2021localizing} (LVS), trained with a localization module and hard negative mining, and for Music Synthesis \cite{hu2020discriminative} (DSOL), trained on music domain data specifically, and require object-level representations. 

\vspace{-0.5em}
\subsection{How does the loss function affect the localization?}
\vspace{-0.5em}

% We are going to take a look of result metrics from each of the model and dataset combination. And going through comparisons of stats of both dataset bounding boxes and predictions from each model. Also take a look of qualatative analysis by looking at several output examples from each dataset and model.

Comparing LVS with RCGrad in Table \ref{tab:results}, we see that the performance of RCGrad is better in FS, relatively close to LVS in VS \cite{chen2021localizing}, and significantly better than the other models in MS and US. RCGrad's consistent performance across datasets and the trends present in Figure \ref{fig:dist} suggest that the hard-negative sampling and learn to localize modules (with size 14x14) might not be able to provide the granularity required for MS and US. For Music Synthesis, RCGrad outperforms SOTA \cite{hu2020discriminative}, which is originally trained on a music dataset with extra object level representations, indicating that training with simple contrastive loss might be sufficient, and generalize better to different domains. An important outcome of our analysis is that the evaluation dataset considerably mediates the conclusions on the usefulness of the different modules of the system for the localization of visual sources, so it is critical to have a diverse set of datasets featuring different conditions and challenges to be able to properly understand them.
% and even beat specialized models. 
% LVS most of the time generates heatmaps around the center region, might indicate that there is overfitting with the combinations of using localization modules and training on VGGSound dataset. 

\vspace{-0.5em}
\subsection{What is the impact of the encoder choice?}
\vspace{-0.5em}
%Comparing ClipAudioGrad with ClipAudioTran in Table \ref{tab:results}, in general Transformer always outperforms Gradient based visualization techniques. This makes sense as CLIP image encoder is a Vision transformer and transformer based approach has been shown better results across the board comparing with Grad-Cam from the original literature \cite{chefer2021generic}. 
When comparing CLIPAudioTran with RCGrad, we are seeing a big gap in performance, especially for more challenging datasets, i.e. MS and US. This is to our surprise since ViT is a more powerful image encoder (w/ 87.8M parameters, compared to 11.7M for ResNet), and maintains spatial information better than ResNet \cite{raghu2021vision}. Also, CLIP is trained with massive amounts of data, and Wav2CLIP shows pretty promising results in most audio classification tasks \cite{wu2021wav2clip}. To understand this difference better, we look at CLIPTextTran, as shown in Table \ref{tab:results}, CLIPTextTran performs well in both VS and US, even better than RCGrad. This result aligns with our intuition that the ViT encoder and the exposure to considerable amounts of data during training should make CLIP-based models more competitive. The fact that RCGrad does better than CLIPAudioTran potentially indicates that there might be semantic mismatch between image-text and image-audio for the distillation from CLIP to Wav2CLIP, which it might be better to train image-text and audio from scratch, or that the distillation from image and text to audio should be done using more data.%and Wav2CLIP might still be underfitted from the distillation process via VGGSound, compared to the amount of data used to train CLIP originally.

\section{Conclusions}
% \vspace{-0.5em}

In this work, we discuss the impact of different design choices and dataset evaluation benchmarks in the adaptability and understanding of localization methods. Furthermore, we study the interaction between these decisions, the model performance, and the data, by digging into different evaluation datasets spanning different difficulties and characteristics, and discuss the implications of such decisions in the context of real-world applications. We also show that it is critical to have evaluation datasets featuring diverse conditions, including location of sources, number of sources and size. For future work, we would like to explore the use of pre-processing and de-noising techniques to help boost the foreground objects versus the background noises in the audio, include temporal context via sequences of images, and include spatial information from multi-channel audio to further distinguish moving from still objects.

%However, the approach is still prone to background noises, and performs bad on very small objects. 
%. 2. Use another modality, such as text labels together to help clean up the predictions. 3. E

% Also, might confused with still object versus moving object, which is important for real-world traffic monitoring systems. We think that there are ways to improve those by 1. de-noising techniques to help boost the foreground object sound versus the background environmental sound (such as wind) 2. Use another modality such as text labels to help clean up the predictions 3. use stereo sound to complement the predictions.

% We also noticed that this approach will tend to attribute to something on the image even if it is a tiny dots, and sometimes cause the performance to drop, some mechanisms to model false positives or sound that is not on the screen is a good direction to pursue.

\bibliographystyle{IEEEtran}

\bibliography{mybib}

% \begin{thebibliography}{9}
% \bibitem[1]{Davis80-COP}
%   S.\ B.\ Davis and P.\ Mermelstein,
%   ``Comparison of parametric representation for monosyllabic word recognition in continuously spoken sentences,''
%   \textit{IEEE Transactions on Acoustics, Speech and Signal Processing}, vol.~28, no.~4, pp.~357--366, 1980.
% \bibitem[2]{Rabiner89-ATO}
%   L.\ R.\ Rabiner,
%   ``A tutorial on hidden Markov models and selected applications in speech recognition,''
%   \textit{Proceedings of the IEEE}, vol.~77, no.~2, pp.~257-286, 1989.
% \bibitem[3]{Hastie09-TEO}
%   T.\ Hastie, R.\ Tibshirani, and J.\ Friedman,
%   \textit{The Elements of Statistical Learning -- Data Mining, Inference, and Prediction}.
%   New York: Springer, 2009.
% \bibitem[4]{YourName17-XXX}
%   F.\ Lastname1, F.\ Lastname2, and F.\ Lastname3,
%   ``Title of your INTERSPEECH 2022 publication,''
%   in \textit{Interspeech 2022 -- 23\textsuperscript{rd} Annual Conference of the International Speech Communication Association, September 18-22, Incheon, Korea, Proceedings, Proceedings}, 2022, pp.~100--104.
% \end{thebibliography}

\end{document}